\documentclass{aastex6}

\usepackage{epstopdf}
\begin{document}
\title{Study of the abundance features of the metal-poor star HD 94028}
\author{Wanqiang Han\altaffilmark{1,2}, Lu Zhang\altaffilmark{3}, Guochao Yang\altaffilmark{4}, Ping Niu\altaffilmark{2}, and Bo Zhang\altaffilmark{1}}
\affil{1 Department of Physics, Hebei Normal University, 20 Nanerhuan Dong Road, Shijiazhuang 050024, China; zhangbo@mail.hebtu.edu.cn}
\affil{2 Department of Physics, Shijiazhuang University, Shijiazhuang 050035, China}
\affil{3 College of Mathematics and Information Science, Hebei Normal University, Shijiazhuang 050024, China}
\affil{4 School of Sciences, Hebei University of Science and Technology, Shijiazhuang, 050018, China}
\begin{abstract}

Recent work has found that the metal-poor star HD 94028 shows interesting abundance features. The s-process material (e.g., Ba, La, Ce, and Pb) and r-process material (e.g., Eu, Os, Pt) are mildly overabundant while the element C is not enhanced. However, the observed supersolar ratio [As/Ge](= $0.99 \pm 0.23$) and subsolar ratio [Se/As](= $-0.16 \pm 0.22$) could not be fitted by the combination of s- and r-process abundances. In this work, adopting the abundance decomposition method, we investigate the astrophysical origins of the elements in this star. We find that the primary process of massive stars dominates the production of the light elements and iron-group elements. The lighter neutron-capture elements from Ge to Te mainly come from the weak r-process. The elements from Ba to Pb are synthesized dominantly by the main s- and main r-process. The observed abundance features of Ge, As, and Se in HD 94028 are mainly produced by the weak r-process, which possesses the features of the supersolar ratio [As/Ge] and subsolar ratio [Se/As]. Because Ge is not synthesized solely by the neutron-capture process, As should be the lightest neutron-capture element. Although the i-process has been suggested in several astrophysical environments, it should be superfluous to explaining the abundances of the lighter neutron-capture elements in HD 94028.

\end{abstract}
\keywords{nuclear reactions, nucleosynthesis, abundances - stars: abundances - stars: individual (HD 94028)}
\section{Introduction} \label{sec:intro}
Elements beyond the iron group elements are mainly produced through the slow (s) and rapid (r) neutron-capture processes. The s-process consists of two subcomponents. The weak s-process produces the light neutron-capture elements below $A = 90$ and takes place in the core He-burning and shell C-burning phases of the massive stars \citep{Woos02,Trava04,Pignatari10,Kappeler11,Pumo12}. The main s-process generally produces the elements in the mass number region $88 < A < 208$ \citep{Arlan99} and takes place in asymptotic giant branch (AGB) stars with low to intermediate mass \citep{Galli98,Busso99,Bisterzo11}. Based on the s-process calculations, \citet{Arlan99} obtained the r-process abundances using the residual method. The r-process also consists of two subcomponents. The weak r-process, which is also called ``lighter element primary process'' \citep{Trava04}, can produce the light neutron-capture elements with mass number $A < 130$ and probably takes place in Type II supernovae (SNe II) with the progenitors of $M \geqslant 10 M_\odot$ \citep{Trava04,Ishimaru05,Qian07,Arcones14,Pignatari16}. The main r-process produces both light and heavy neutron-capture elements. The exact site (or sites) of the main r-process is (or are) still not identified \citep{Sned08,Thielemann11}. Low-mass SNe \citep{Whee98,Trava99,Qian07,Bisterzo17} and the neutron star mergers (NSMs) \citep{Lattimer74,Eichler89,Freibur99} have received attention for many years. Based on calculations of the Galactic chemical evolution, \citet{Matteucci14} found that NSMs could account for the entire Eu production in the Galaxy if the neutron star binaries merge within 1 Myr. According to the abundance correlations of the metal-poor stars, \citet{Tsujimoto14} proposed that the light r-process elements are synthesized in the core-collapse SNe and the heavy r-process elements are synthesized in NSMs. Through analysis of the abundance Ba and Eu in the extremely metal-poor stars, \citet{Komiya14} reported that both core-collapse SNe and NSMs could produce the heavy r-process elements in the early Galaxy. \citet{Ramirez15} reported that an NSM is a suitable site for the main r-process elements and that the observed scatter of Eu in the metal-poor stars can be explained by the NSM events. Recently, by means of the chemical evolution model, \citet{Komiya16} studied the r-process abundance produced by NSMs and found that the scatter of the r-process abundance in the metal-poor stars can be reproduced. \citet{Ji2016} studied the abundance characteristics of the r-process-enhanced stars in the ultrafaint dwarf galaxy Reticulum II and suggested that the yield and rate of the r-process are consistent with NSM events rather than core-collapse SN events.

The elemental abundances of the metal-poor stars can provide us with useful information to explore the astrophysical origins of the elements in the early Galaxy. Previous studies of the metal-poor stars indicate that they were born in the interstellar medium polluted by multiple nucleosynthetic events \citep{Burris00,Qian01,2006APJ...653...1145K,Masse10}. Because the metal-poor stars CS 22892-052 and CS 31082-001 are highly enhanced in the heavy r-process elements (heavier than Ba) and show a typical r-process abundance pattern, they are often referred to as the main r-process stars \citep{Sned08}. On the other hand, the metal-poor stars HD 122563 and HD 88609 show an excess of the light neutron-capture elements compared to the heavy ones and are considered as the weak r-process stars \citep{Honda07,Montes07}. Because the light elements and iron-group elements in the main r-process stars did not originate from the r-process and the main r-process elements in the weak r-process stars did not come from the weak r-process, \citet{2013PASP...125...143L} and \citet{Hansen14} extracted the abundances of pure main r- and weak r-process components from the abundances of the main r-process stars and weak r-process stars using an iterative approach. HD 94028 is a metal-poor star with metallicity [Fe/H] =$ -1.62 \pm 0.09$. Using the high-resolution spectra, \citet{2011APJ...742...21P} derived the elemental abundances of HD 94028 and reported that the element Mo was extremely enhanced with an abundance ratio [Mo/Fe] =$ +1.0$. On the basis of the archival space-based ultraviolet spectra, \citet{2012APJ...756...36R} derived the abundances of Ge, As, and Se in HD 94028 and reported that their origins might not be the main s-process or weak s-process. However, \citet{2014AJ...147...136R} derived the abundances of 313 metal-poor stars and found that s-process material is moderately enriched in HD 94028. Recently, \citet{2016APJ...821...37R} analyzed the abundance pattern of HD 94028 and found that the star exhibits moderate enhancement of s-process material and r-process material, but that the observed abundances, especially for the supersolar ratio [As/Ge](= $0.99 \pm 0.23$) and subsolar ratio [Se/As](= $-0.16 \pm 0.22$), could not be explained by the s-process abundances (weak s- or main s-), the r-process abundances (weak r- or main r-), or the combination of s- and r-process abundances. In order to explain the abundance pattern of HD 94028, they suggested that another additional process, i.e., the intermediate neutron-capture process (i-process), was required. Up to now, the astrophysical site (or sites) of the i-process has (or have) not been confirmed \citep{2016APJ...821...37R}. Observationally, both the main s-process and the i-process are closely associated with the AGB stars \citep{Camp10,Herw11,Lugaro2015}. Obviously, a detailed study of the abundances is still needed to investigate the astrophysical origins of the elements in HD 94028.

The Elements Ge ($Z = 32$), As ($Z = 33$), and Se ($Z = 34$) lie in between the iron peak and the neutron-capture elements.  The proton-capture process (p-process) occurs in the O/Ne-rich layers of massive stars during pre-explosion or supernova phase and produces the heavy neutron-deficient nuclides \citep{Arnould03}.
Based on the calculations of the p-process nucleosynthesis, \citet{Rayet90} found that the p-process is not the dominant transformation path for nuclides with $Z\geqslant32$. Through analyzing the abundance relations of the neutron-capture elements, including Ge, in the metal-poor stars with $-3.1 \leqslant$ [Fe/H] $\leqslant -1.6$, \citet{Cowan05} found that the abundance patterns for the heavier neutron-capture elements match the r-process pattern in the solar system well, which implies that the heavier neutron-capture elements in the metal-poor stars possess a common astrophysical origin. Furthermore, they found that there exists a correlation between Ge abundances and Fe abundances: [Ge/H] =[Fe/H]$-0.79$, which could imply that the charged-particle or explosive synthesis was the origin of Ge in the metal-poor stars. By comparing the abundances of the metal-poor stars with the yields of the massive stars, \citet{2014MNRAS...443...2426N} found that the observed Ge abundances are higher than the scaled Ge abundances of the primary-like yields by about 1.0 dex, and they suggested that Ge mainly originates from the r-process for the early Galaxy. Furthermore, they found that, for the solar system, about 59\% of Ge originates from the neutron-capture process and about 41\% of Ge originates from the secondary-like yields of massive stars.

In this work, based on the component abundances of Ge calculated by \citet{2014MNRAS...443...2426N} and the abundance method presented by \citet{Li2013}, we study the elemental abundances in HD 94028. The model and calculations are presented in section 2. The discussion of the abundance features is contained in section 3, and our conclusions are given in section 4.

\section{Model and Calculations} \label{sec:calcu}
The chemical elements in the metal-poor stars should originate from multiple nucleosynthetic mechanisms \citep{Trava99,Trava04,2010APJ...714...L123R}. In order to explain the r-process abundances of the metal-poor stars, \citet{Qian01} suggested that there are two nucleosynthetic events: the high-frequency events (H events) and the low-frequency events (L events). The H events mainly produce heavy r-elements ($A>130$) and the L events mainly produce Fe and light r-elements ($A\leq130$).
By assuming that only two nucleosynthesis processes (the H and L events) dominate the production of the neutron-capture elements in the metal-poor stars, \citet{Hansen14} found that the abundance patterns of the two processes are robust within the attributed uncertainties, and that the abundance patterns of most metal-poor stars can be successfully fitted by a linear superposition of the two processes. Our primary goal is to analyze the astrophysical origins of elements from O to Pb in HD 94028 and study the abundance features of this star. The abundance of the ith element in HD 94028 can be decomposed according to the following five-parametric model \citep{Li2013}:
\begin{equation}\label{1}
  N_i=(C_{r,m}N_{i,r,m}+C_{pri}N_{i,pri}+C_{s,m}N_{i,s,m}+C_{sec}N_{i,sec}+C_{ia}N_{i,Ia})\times10^{[Fe/H]},
\end{equation}
where $N_{i,r,m}$, $N_{i,pri}$, $N_{i,s,m}$, $N_{i,sec}$ and $N_{i,Ia}$ represent the scaled component abundances of the main r-, primary, main s-, and secondary processes, and SNe Ia, respectively. $C_{r,m}$, $C_{pri}$, $C_{s,m}$, $C_{sec}$ and $C_{Ia}$ are the superposition coefficients. Using the equation (1) we can calculate the abundance originating from each process. The scaled abundances of the SNe Ia are taken from \citet{1995APJ...98...617T}, in which the scaled abundances of Fe, Ni, Cu, and Zn are updated from the SNe Ia yields presented by \citet{Ohkubo06}. The component abundances of the main r- and primary processes are taken from \citet{2013PASP...125...143L}. The primary elements consist of the primary light elements and weak r-process elements, since these elements are ejected from massive stars. The Ge component abundances are adopted from \citet{2014MNRAS...443...2426N}.

Based on an analysis of the metallicity-dependent yields of the low-mass AGB stars, \citet{Busso01} suggested that the ratio [Pb/hs] would be larger than 1.0 with [Fe/H] $\leqslant-1.3$, where hs represents the heavy s-process elements. Three lead stars (HD 187861, HD 224959 and HD 196944) have been found \citep{Van01}. Meanwhile, two s-enhanced metal-poor stars (LP 706-7 and LP 625-44) with [Pb/Ce] $< 0.4$, which cannot be considered as lead stars, have been reported \citep{Aoki01}. Observationally, the ratio [Pb/hs] for the s-rich metal-poor stars shows a large spread \citep{Aoki02,Cohen03,Lucatello03,Van03,Johnson02,04,Sivarani04}. \citet{Straniero06} suggested that a large spread of $^{13}C$-pocket efficiencies is needed to account for the observed spreads of [Pb/hs] in s-rich and lead-rich stars. \citet{2010MNRAS...404...1529B} calculated the abundances of the low-mass metal-poor AGB stars for different $^{13}C$-pocket efficiencies, including the standard case (hereafter ST), ST/12, ST/45, and ST/75. In the ST case \citep{Galli98}, low-mass AGB stars with a half metallicity of the solar system can exactly reproduce the main s-process abundances of the solar system \citep{Arlan99}. On the basis of the abundance analysis of HD 94028, \citet{2016APJ...821...37R} found that the elements Ba and Pb are enhanced, which means that the natal cloud in which the star formed is polluted by the s-process material of low-mass AGB stars with low metallicity. In equation (1), the scaled abundances of $N_{i,s,m}$ are adopted from the s-process abundances of the 2.0 $M_\odot$ AGB model at [Fe/H] $= -2.6$ presented by \citet{2010MNRAS...404...1529B} for case ST/18, since the ratio [Pb/Ba]$= 0.44$ of this case is close to the observed ratio [Pb/Ba]$= 0.47$. The C abundance is not included in our calculations, since the predictions of the ratio [C/Fe] are largely overestimated by AGB models \citep{Bisterzo11}. The lifetime of a 2.0 $M_\odot$ AGB star is about 1.2 Gyr \citep{Karakas14}. Because of the delay of the stellar lifetimes, the AGB star could pollute the natal cloud with metallicity [Fe/H]= -1.6, which is close to the metallicity of HD 94028.

The secondary elements consist of the secondary light elements and weak s-process elements, since these elements originate from massive stars \citep{Li2013}. The component abundances of the secondary process are taken from \citet{2013PASP...125...143L}. Considering the secondary nature of the neutron source $^{22}Ne(\alpha, n)^{25}Mg$, the yields of the weak s-process decrease approximately linearly with the decreasing metallicity \citep{Mathews92,Bisterzo17}. \citet{Trava04} suggested that the contribution of the weak s-process is very little for halo stars, and about 10\% of the Sr abundance comes from the weak s-process in the solar system. However, considering the effect of rotation-induced mixing on the
weak s-process, \citet{Frischknecht12} calculated the weak s-process yields for a massive 25 $M_\odot$ massive star and reported that the primary production of $^{22}Ne$ should occur in fast-rotating massive stars at very low metallicity and the weak s-process could contribute significantly to the production of elements up to Ba. Recently, \citet{Frischknecht16} calculated the weak s-process yields for rotating and non-rotating massive stars with four metallicities ($Z = Z_\odot$, $10^{-3}$, $10^{-5}$, and $10^{-7}$). The calculated results imply that the yields of the weak s-process depend on the initial metallicity of the massive stars. In this work, the abundances of the weak s-process are adopted from two cases.

(i) Case A. Because of the secondary-like nature of the neutron source, the abundances of the weak s-process at solar metallicity should play an important role \citep{Trava04}. The abundances of the weak s-process are taken from \citet{Raiteri93}, which presents the abundances of the weak s-process at the epoch of formation of the solar system.

(ii) Case B. Considering the metallicity-dependent yields, the abundances of the weak s-process of a non-rotating massive star with $Z = 10^{-3}$ calculated by \citet{Frischknecht16} are adopted, since the metallicity [Fe/H] of HD 94028 is about -1.62.

Based on a study of the chemical compositions of extremely metal poor stars, \citet{Ishimaru05} stated that the weak r-process yields are large for the light neutron-capture elements but decrease with increasing atomic mass. \citet{Honda06} found that the abundances of the neutron-capture elements for the weak r-process star (i.e. HD 122563) decrease gradually with increasing atomic number, and they concluded that the decreasing trend is a key to investigating the characteristics of the weak r-process. Based on a comparison of the abundances of the weak r-process star with the abundances of the r-process in the solar system, \citet{Montes07} found that the relative contributions of the weak r-process to the solar system abundances decrease with increasing atomic number. Using the iterative method, \citet{2013PASP...125...143L} derived the two r-process abundances from the abundances of the weak r-process stars and the main r-process stars. They defined the `percentage of weak r-process component' $f^{r}_{r,w}$ (i.e., $N_{i,r,w}/(N_{i,r,m} + N_{i,r,w})$) to express the relative contributions of the weak r-process to the r-process abundances of the solar system and found that $f^{r}_{r,w}$ decreases linearly with increasing atomic number. For neutron-capture elements, the average deviation of the contributed fractions from the linear relation is about 9\%. For the element Ge, \citet{2014MNRAS...443...2426N} calculated the percentages contributed by the main r-process and the weak r-process and reported that $f^{r}_{r,w}$ is about 91\%. This result is in agreement with the linear relation and the deviation is about 4\%. Furthermore, \citet{Hansen14} and \cite{Niu15} found that the abundance patterns and the decreasing trends of the weak r-process abundances vs. the atomic number are robust. In this work, using the decreasing relation, we estimate the abundances of the weak r- and main r-process for the elements As and Se, which are listed in Table 1. For comparison, the corresponding r-process abundances in the solar system \citep{Sned08} are also listed.

\setlength{\tabcolsep}{12pt}
\begin{table}[h]
\begin{center}
\caption{The main r-process and weak r-process abundances of As, Se adopted in this work.}
\label{tab1}
\begin{tabular}{lcccc}
\hline\noalign{\smallskip}
Element & Z & N$_{i,r,tot}$ & N$_{i,r,m}$ & N$_{i,r,w}$ \\
\noalign{\smallskip}
\hline
\noalign{\smallskip}
As  & 33 & 5.33\tablenotemark{a} & 0.43 & 4.90 \\
Se  & 34 & 40.26\tablenotemark{a} & 4.43 & 35.83 \\
\hline
$^a$\citet{Sned08}.\\
\end{tabular}
\end{center}
\end{table}
\setlength{\tabcolsep}{2.0pt}

The five superposition coefficients in Equation (1) can be derived by comparing the calculated abundance $N_{i,cal}$ with the observed abundance $N_{i,obs}$ and seeking the smallest $\chi^2$ which is calculated as

 \begin{equation}\label{2}
   \chi^2=\sum{^K_{i=1}\frac{(\log{N_{i,obs}}-\log{N_{i,cal}})^2}{(\triangle\log{N_{i,obs}})^2(K-K_{free})}},
 \end{equation}
where $\Delta\log{N_{i,obs}}$ is the observed uncertainty. Since thirty-eight elements are used in the calculations, thus $K=38$. The value of $K_{free}$ is five, which means five superposition coefficients. All observed abundances of the star HD 94028 are taken from \citet{2016APJ...821...37R}. The abundances derived from neutral lines are adopted for the light neutron-capture elements Mo and Cd.

For the best-fitting results, the value of $\chi^2$ is 0.887 for case A and 0.938 for case B, respectively. The calculated superposition coefficients for case A are $C_{r,m} = 1.50$, $C_{pri} = 3.90$, $C_{s,m} = 1.84$, $C_{sec} = 0.17$, and $C_{Ia} = 0$, respectively. For case B, these coefficients are $C_{r,m} =1.49$, $C_{pri} =4.10$, $C_{s,m} =1.82$, $C_{sec} =0.13$, and $C_{Ia} =0$, respectively. The coefficients $C_{r,m}$, $C_{pri} $, and $C_{s,m}$ are larger than 1.0, which means that the fractions contributed by the main r-, primary, and main s-process for HD 94028 are larger than those of the solar system. On the other hand, $C_{sec}$ is smaller than 1.0, which implies that the fraction contributed by the secondary process for this star is smaller than that of the solar system. The coefficient $C_{Ia}$ is 0, which is consistent with the prediction of \citet{Kobay98} that the SNe Ia did not occur for the progenitors with [Fe/H] $\leqslant -1.0$.

\begin{figure}[h]
\includegraphics[width=3.5in,height=2.8in]{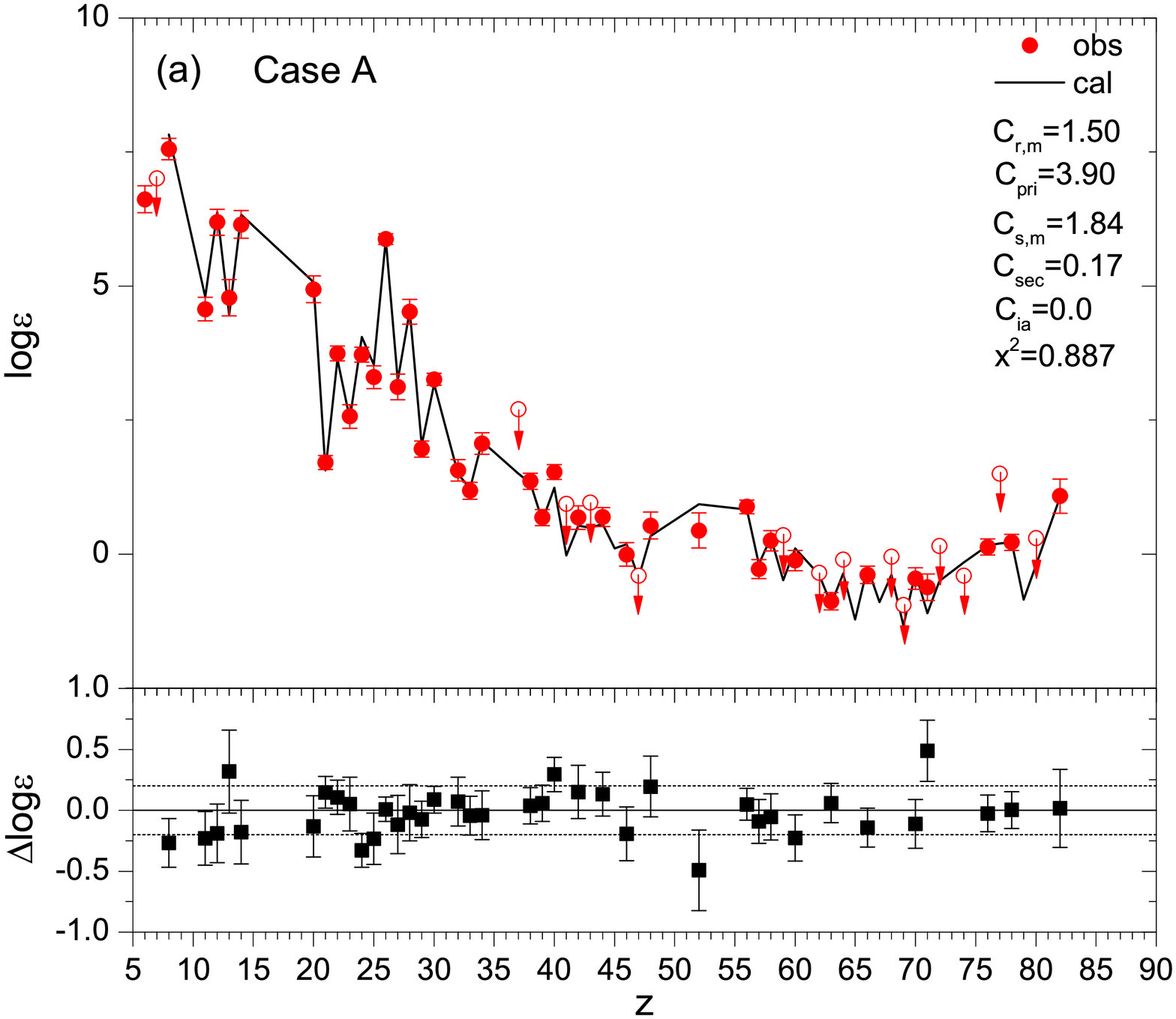}
\includegraphics[width=3.5in,height=2.8in]{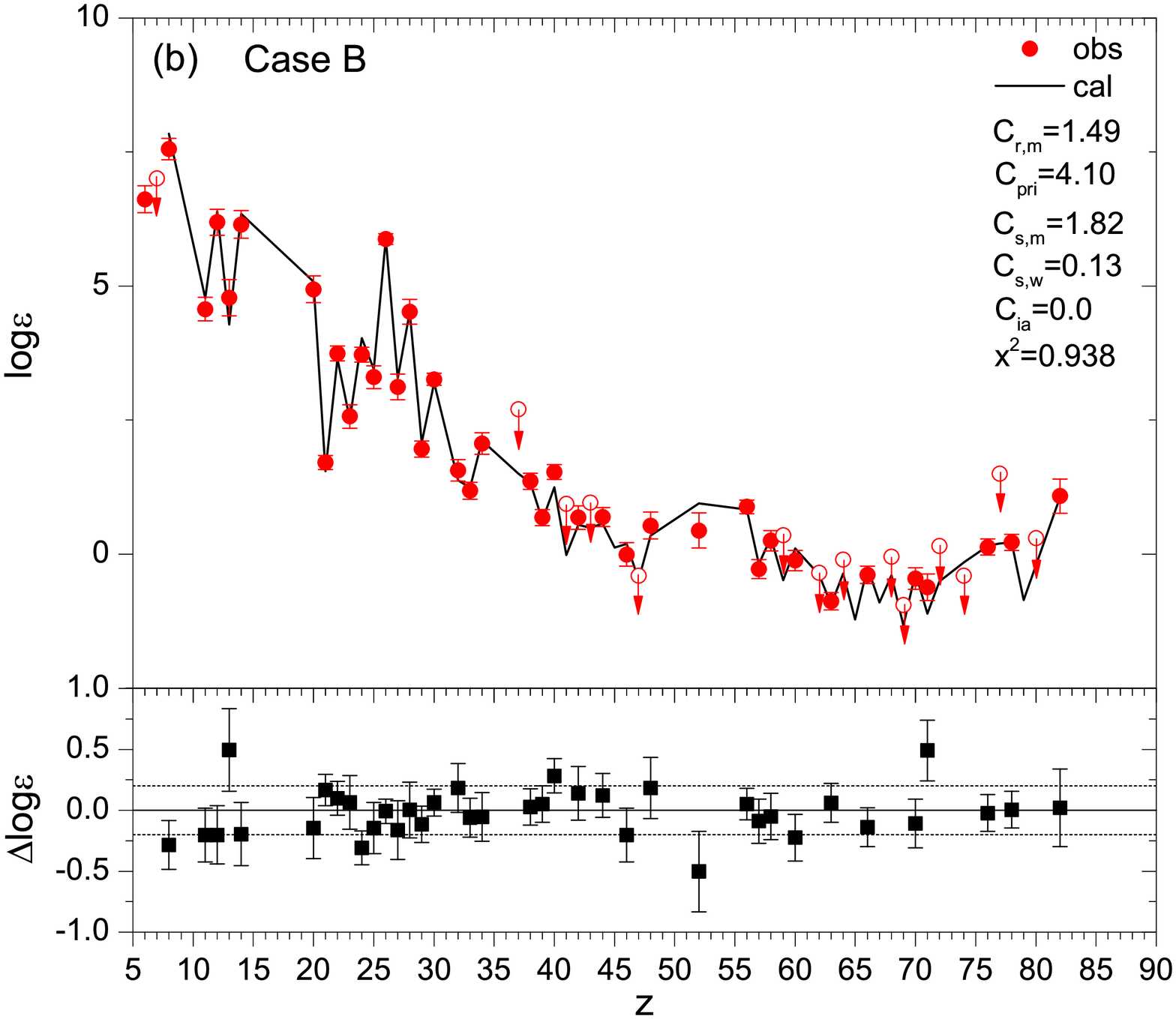}
\caption{Upper: The comparisons of calculated results with the observations. The observed abundances are represented by filled circles and the calculated abundances are marked as the solid lines. The observed upper limits are represented by circles with downward arrows. The data are taken from \citet{2016APJ...821...37R}. Lower: the relative offsets ($\Delta\log{\varepsilon}=\log{N_{i,obs}}-\log{N_{i,cal}}$ ) and the observed uncertainties.\label{f1}}
\end{figure}

Direct comparisons of calculated and observed abundances for case A and case B are plotted in Figures 1(a) and (b), respectively. In the upper panels, the solid lines are the calculated abundances and the filled circles are the observed abundances of HD 94028. The lower panels show the related offsets ($\Delta\log{\varepsilon}=\log{N_{i,obs}}-\log{N_{i,cal}}$)
and the corresponding observed uncertainties. From Figures 1(a) and (b), we can see that, for the most elements, the calculations are consistent with the observations within the uncertainties. Comparisons between the individual component abundances and the observed abundances in HD 94028 are shown in Figures 2(a) and (b).

\begin{figure}[h]
\includegraphics[width=3.5in,height=2.8in]{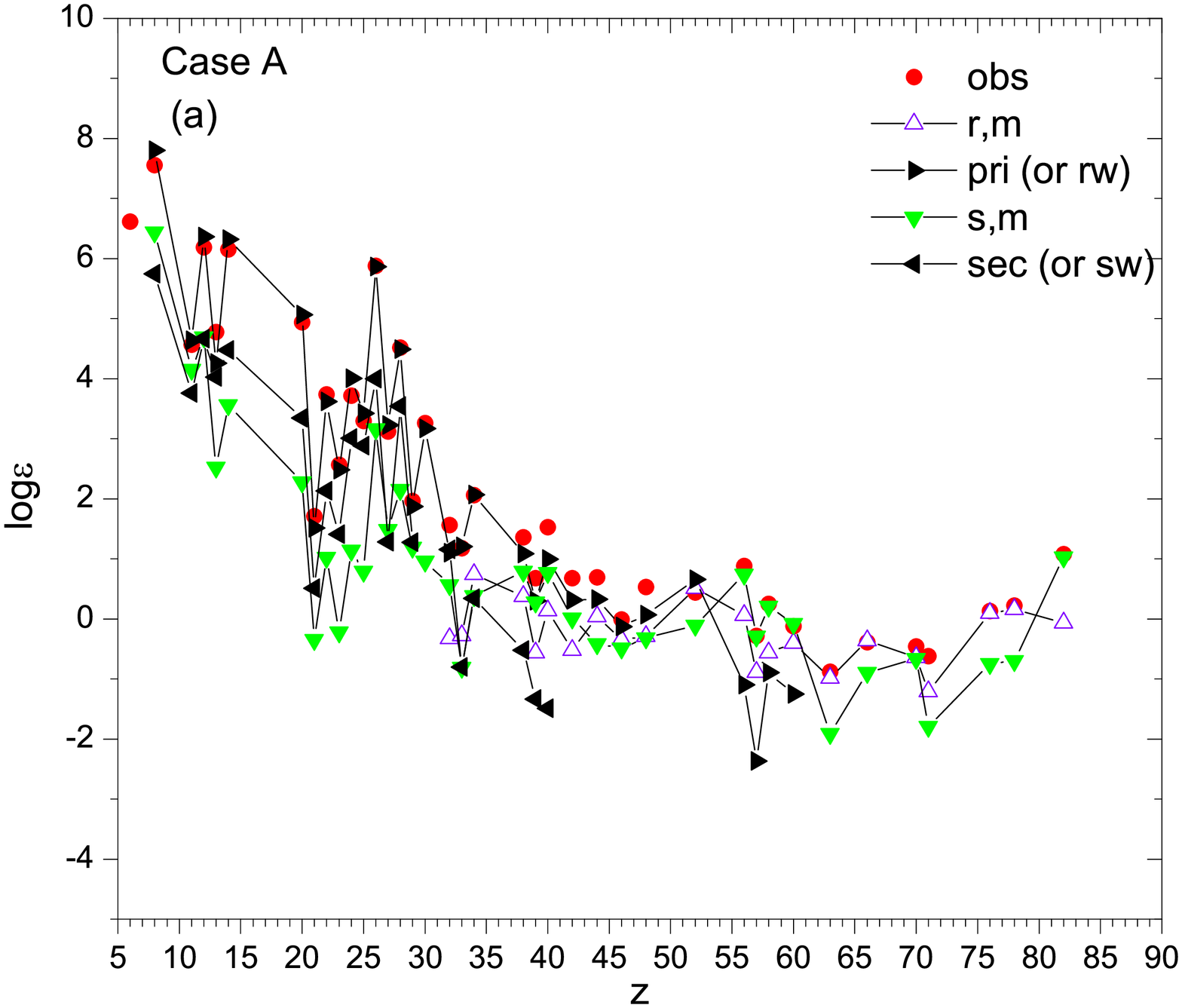}
\includegraphics[width=3.5in,height=2.8in]{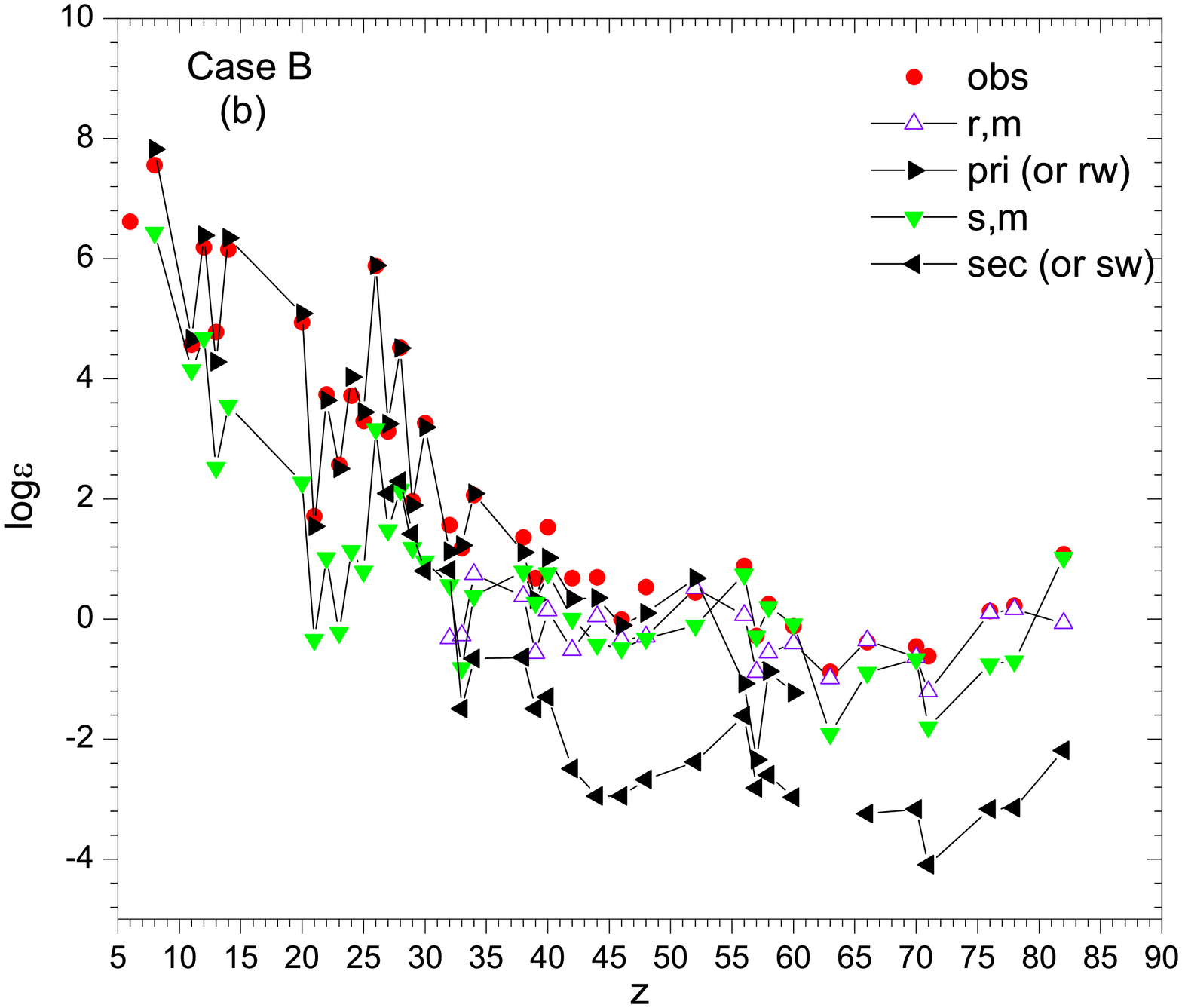}
\caption{The abundance patterns of various components in HD 94028. The patterns of main r-, primary, main s- and secondary components are marked by triangle, filled right triangle, filled down triangle and filled left triangle, respectively. Filled circles represent the observed abundances which are taken from \citet{2016APJ...821...37R}. \label{f2}}
\end{figure}

\section{Discussion} \label{sec:discuss}

In order to study the astrophysical origins of each element in the sample star, Figures 3(a) and (b) show the calculated abundance ratios and components ratios. For comparison, the observed abundance ratios of HD 94028 are also shown. The left panels show that the origin of the light elements from O to Zn is dominantly ascribed to the primary process of the massive stars. For these elements, the contribution of the secondary process can be negligible, as the secondary yields of the massive stars decrease with decreasing metallicity. The right panels show that the origin of the heavy neutron-capture elements Ba, La, Ce, Nd, and Pb is dominantly ascribed to the low-mass metal-poor AGB stars. Furthermore, the origin of Eu, Dy, Lu, Os, and Pt is mainly ascribed to the main r-process. The astrophysical origins of the heavy neutron-capture elements from Eu to Pt have been analyzed by \citet{2016APJ...821...37R}. They concluded that the heavy rare-earth elements and third r-process peak mainly originate from the r-process. For the astrophysical origins of the heavy neutron-capture elements from Eu to Pt, our result is in agreement with their conclusion.

\begin{figure}[h]
\includegraphics[width=6.5in,height=2.8in]{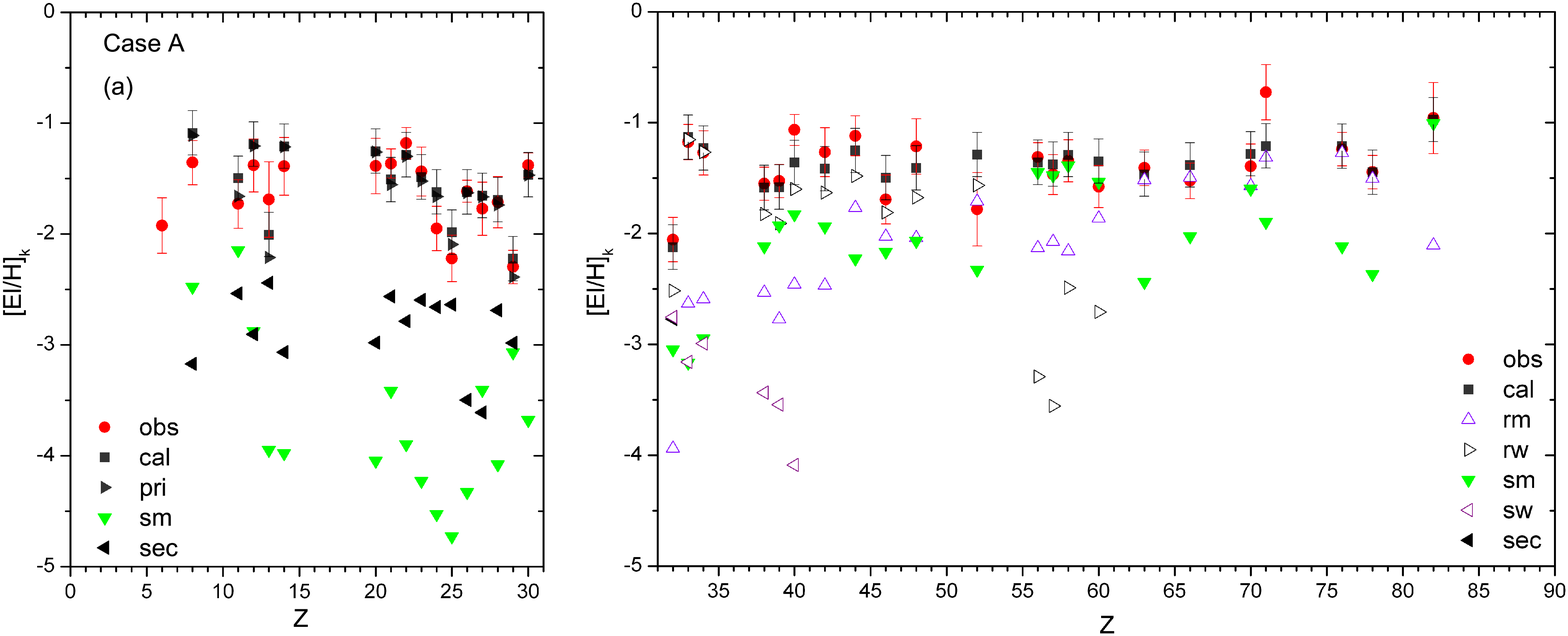}
\includegraphics[width=6.5in,height=2.8in]{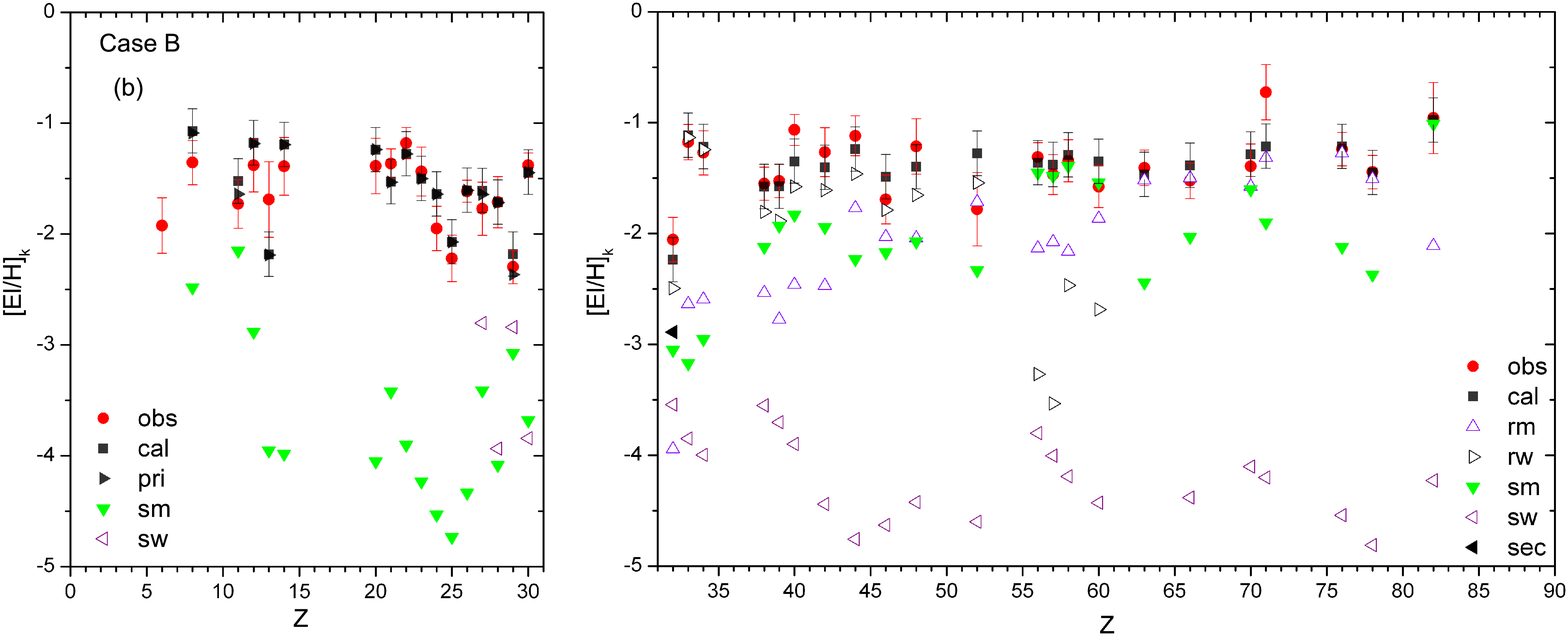}
\caption{Component ratios $[X_k/H]$ of the elements of HD 94028. The filled circles are the observed abundance ratios and the filled squares represent the calculated ratios. The filled right triangles, filled left triangles, open triangles, open right triangles, filled down triangles and open left triangles are the ratios of primary, secondary, main r-, weak r-, main s- and weak s-components, respectively. \label{f3}}
\end{figure}


Based on the abundance analysis of HD 94028, \citet{2016APJ...821...37R} reported that the abundance ratio [Ba/Fe] is 0.32 and the ratio [C/Fe]($=-0.06\pm0.19$) is approximately equal to that of the solar system. The predicted s-process abundance ratio is [Ba/Fe]$=2.88$ and the C abundance ratio is [C/Fe]$=4.06$ for a 2.0$M_\odot$ AGB star with [Fe/H] $= -2.6$ \citep{2010MNRAS...404...1529B}. Assuming that the elements Ba and C in HD 94028 mainly originate from the metal-poor AGB star, we estimate the dilution factor of the s-process material of the AGB star. Firstly, the ratios [Ba/Fe] and [C/Fe] of the s-process material of the AGB star should be decrease by about 1.0 dex, since the metallicity of HD 94028 is higher than that of the AGB star by about 1.0 dex. Secondly, the ratio [Ba/Fe] of the s-process material decreased from 1.88 to 0.32 due to the dilution of the interstellar gas in which the star formed. In this case, the dilution factor of the s-process material of the AGB star is about 2.7\%. The ratio [C/Fe] of the s-process material decreased to the solar ratio also due to the dilution of the interstellar gas. Adopting the observed upper limit of the ratio [C/Fe] as the constraint, the upper limit of the ratio [C/Fe] of the AGB star should be about 2.69, which is lower than the predicted ratio [C/Fe] about 1.37 dex. This result confirms the suggestion that [C/Fe] is largely overestimated by AGB models \citep{Bisterzo11}. Obviously, the observed low ratio [Ba/Fe] is due to the small dilution factor for the s-process material of the AGB star and this small dilution factor should be the astrophysical reason for the observed low ratio [C/Fe] in HD 94028.

\citet{2012APJ...756...36R} studied abundance trends of the ratios [Ge, As, Se/Fe] as a function of [Fe/H] in the metal-poor stars and found that the ratio [Ge/Fe] increases with increasing [Fe/H]. However, the ratio [As/Fe] shows a flattened trend with increasing [Fe/H]. The increasing trend of [Ge/Fe] implies that, apart from the neutron-capture process, Ge has an additional astrophysical origin. \citet{2014MNRAS...443...2426N} studied the astrophysical origins of Ge element and found that it is not a `pure' neutron-capture element. The increasing trend of [Ge/Fe] is partly due to the contribution of the secondary-like yields of the massive stars. Recently, \citet{2016APJ...821...37R} found that, compared to the solar system, the abundance ratio [As/Ge](=$0.99\pm0.23$) is high but the ratio [Se/As](=$-0.16\pm0.22$) is low in HD 94028. In this work, adopting the r-process abundances of Ge presented by \citet{2014MNRAS...443...2426N} and the r-process abundances of As and Se listed in Table 1, we can see from the right panels of Figures 3(a) and (b) we can see that the Ge, As, and Se abundances in HD 94028 can be fitted within the observed uncertainties. For this star, the Ge abundances of the weak r-process are higher than those of the weak s-process and the secondary process. On the other hand, the panels show that the elements As and Se in HD 94028 dominantly come from the weak r-process. Note that the fitted results imply that the elements As and Se are the `pure' neutron-capture elements. The observed flattened trend of [As/Fe] for the metal-poor stars also implies a constant As/Fe ratio for the ejecta of the SNe II. Because Fe dominantly comes from the primary-like yields of the massive stars, As should also be mainly ejected from the SNe II.

For the metal-poor stars, the average ratios of [As/Ge] and [Se/As] are about 0.9 and -0.1, respectively \citep{2012APJ...756...36R}. The ratio [As/Ge] of the r-process in the solar system is close to 0.18 \citep{Sned08} and the ratio [As/Ge] is smaller than 0.0 for the AGB stars \citep{2016APJ...821...37R}. Because the ratios [As/Ge] of
the r-process and s-process are lower than 0.2, \citet{2016APJ...821...37R} reported that all the calculated abundance ratios [As/Ge] with the combination of r-process and s-process yields are lower than the observed [As/Ge](=$0.99\pm0.23$) in HD 94028. They suggested that the supersolar ratio [As/Ge] and the subsolar ratio [Se/As] are common features for the metal-poor stars. In this case, in order to fit the elemental abundances, an additional contribution from the i-process is suggested in order to fit the elemental abundances. They reported that, for the i-process, the abundance ratio [As/Ge] is supersolar and the ratio [Se/As] is subsolar, which is consistent with the abundance features of the metal-poor stars.

For HD 94028, the measured ratios [As/Ge] and [Se/As] are $0.99\pm0.23$ and $-0.16\pm0.22$. The calculated ratios [As/Ge] and [Se/As] are 0.99 and -0.10 for case A (and 1.12 and -0.10 for case B). In order to explain the abundance features of Ge, As, and Se in detail, we define the component ratios of each process as
\begin{equation}\label{4}
   [E_i/E_j]_k= \log{(N_{i,k}/N_{j,k})}-\log{(N_{i,\odot}/N_{j,\odot})},
\end{equation}
where k=r,m; r,w; s,m; s,w.
Adopting the Ge abundances of the weak r-process and main r-process presented by \citet{2014MNRAS...443...2426N} and the As and Se abundances of the weak r-process and main r-process listed in Table 1, the calculated component ratios of [As/Ge]$_{r,w}$, [As/Ge]$_{r,m}$, [Se/As]$_{r,w}$ and [Se/As]$_{r,m}$ are about 1.36 , 1.31, -0.11, and 0.04, respectively. On the other hand, the component ratios of [As/Ge]$_{s,m}$ and [Se/As]$_{s,m}$ are about -0.12 and 0.22 \citep{2010MNRAS...404...1529B}. The ratios of [As/Ge]$_{s,w}$ and [Se/As]$_{s,w}$ are -0.4 and 0.17 for case A \citep{Raiteri93}, and the weak s-component ratios are -0.31 and -0.15 for case B \citep{Frischknecht16}. It is important to note that it is only for the weak r-process that the component ratio [As/Ge]$_{r,w}$ is supersolar and the ratio [Se/As]$_{r,w}$ is subsolar.
Recalling that the observed ratio [As/Ge] is super-solar and the ratio [Se/As] is sub-solar, the weak r-process should be responsible for the observed abundance features of Ge, As, and Se in HD 94028.
\begin{figure}[h]
\includegraphics[width=3.5in,height=2.8in]{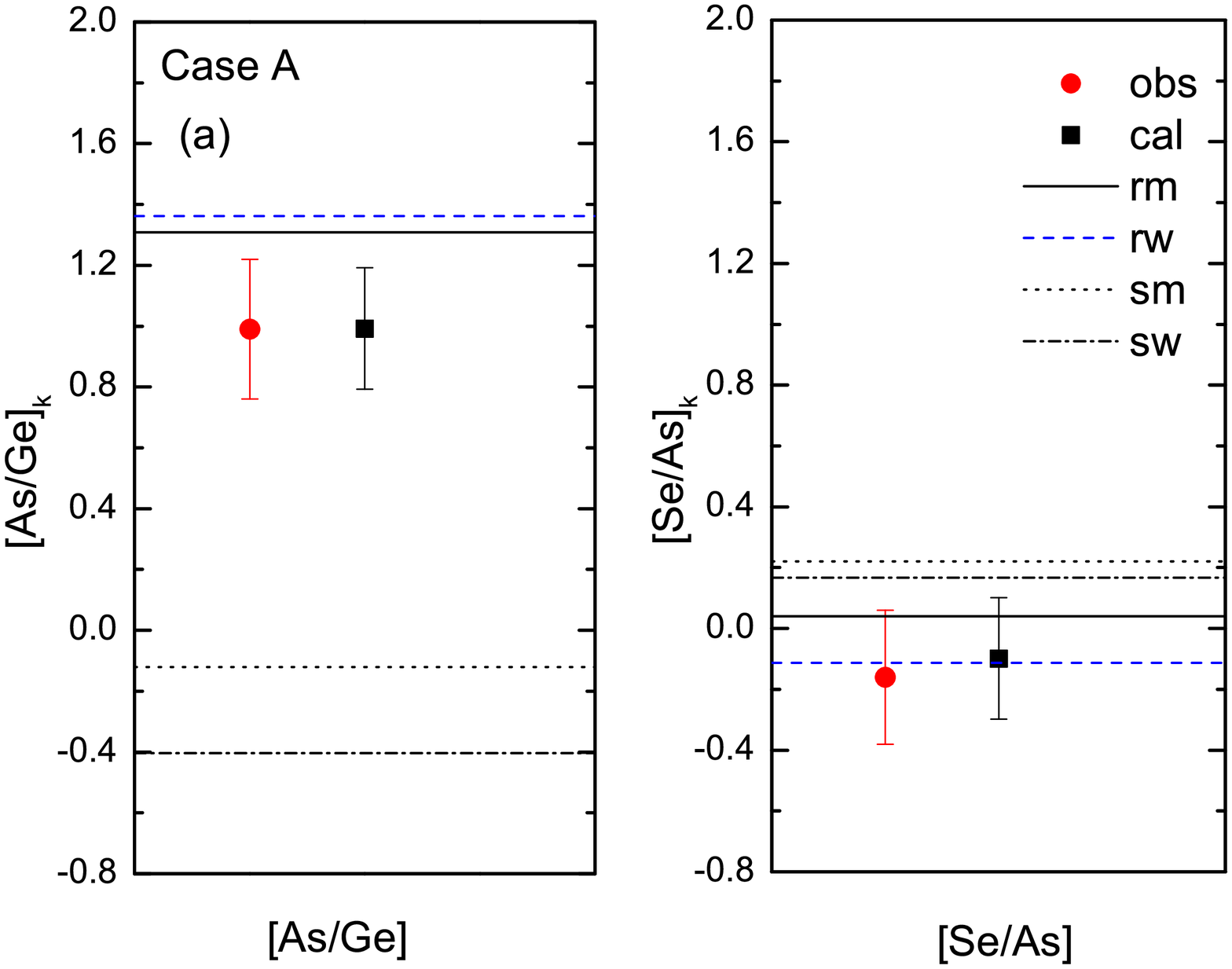}
\includegraphics[width=3.5in,height=2.8in]{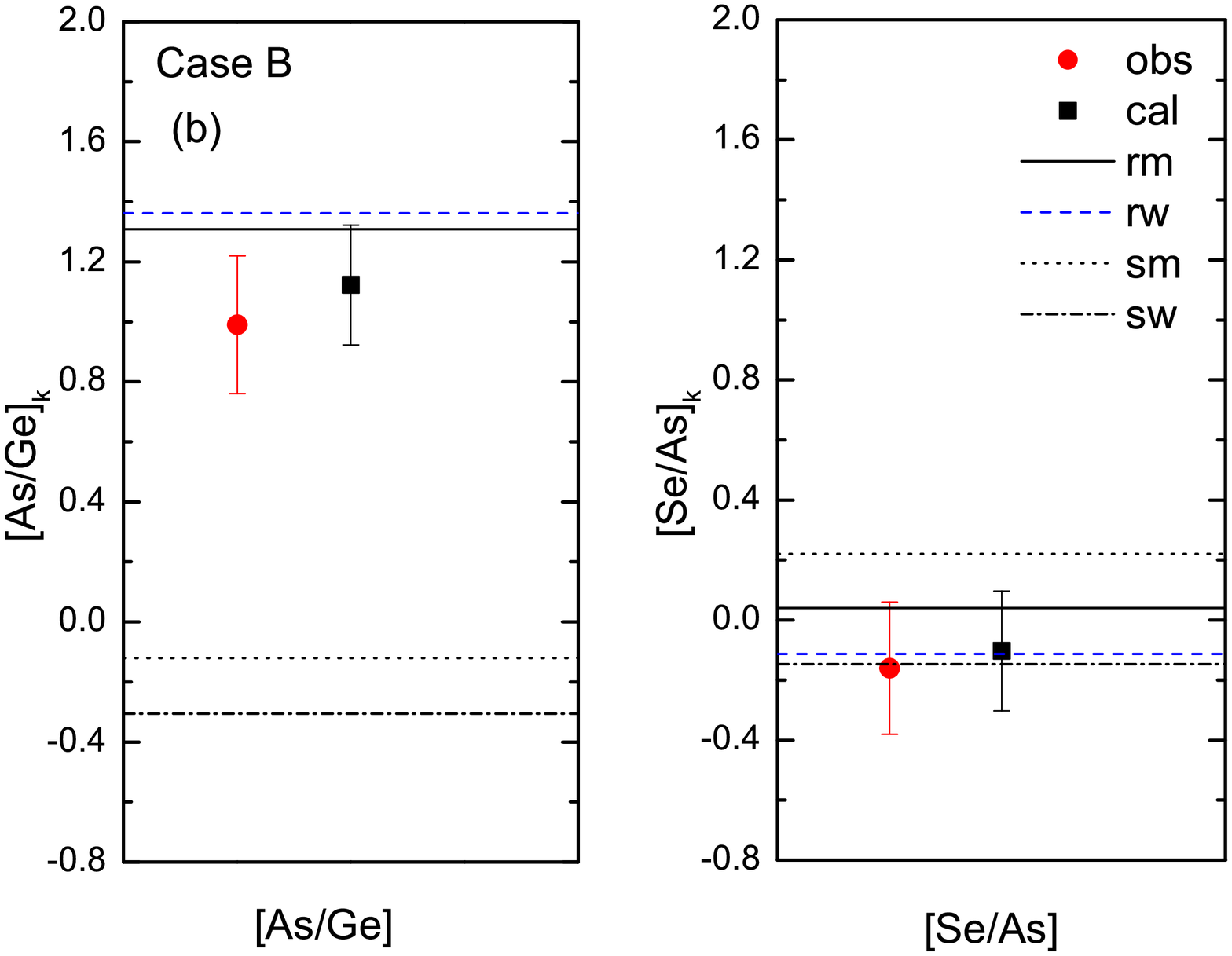}
\caption{The comparisons of the component ratios [As/Ge]$_k$ and [Se/As]$_k$ of HD 94028. The filled circles are the observed ratios, and filled squares are calculated ratios. The solid lines, dashed lines, dotted lines and dash-dotted lines represent the contributions of main r-, weak r-, main s- and weak s-components, respectively.   \label{f4}}
\end{figure}
Figures 4(a) and (b) show the component ratios [As/Ge]$_k$ and [Se/As]$_k$, respectively. The filled circles are the observed ratios and the filled squares are the calculated ratios. The solid lines, dashed lines, dotted lines, and dashed dotted lines represent the component ratios of main r-, weak r-, main s-, and weak s-components, respectively. From the left panels of Figures 4(a) and (b), we can see that the observed ratio [As/Ge] is higher than [As/Ge]$_{s,m}$ and lower than [As/Ge]$_{r,w}$ but close to the ratio [As/Ge]$_{r,w}$. From the right panels we can see that the observed ratio [Se/As] is also close to the weak r-process ratio [Se/As]$_{r,w}$. Obviously, the supersolar ratio [As/Ge] and the subsolar ratio [Se/As] of HD 94028 are mainly due to the contribution of the weak r-process. Although the initial metallicity affects the abundances of the weak s-process, from Figures 1-4 we can see that the difference in the calculated results between case A and case B is small, since the contribution of the weak s-process to the lighter neutron-capture elements of HD 94028 is smaller than that of the weak r-process.
\begin{figure}[h]
\centering
\includegraphics[width=3.5in,height=2.8in]{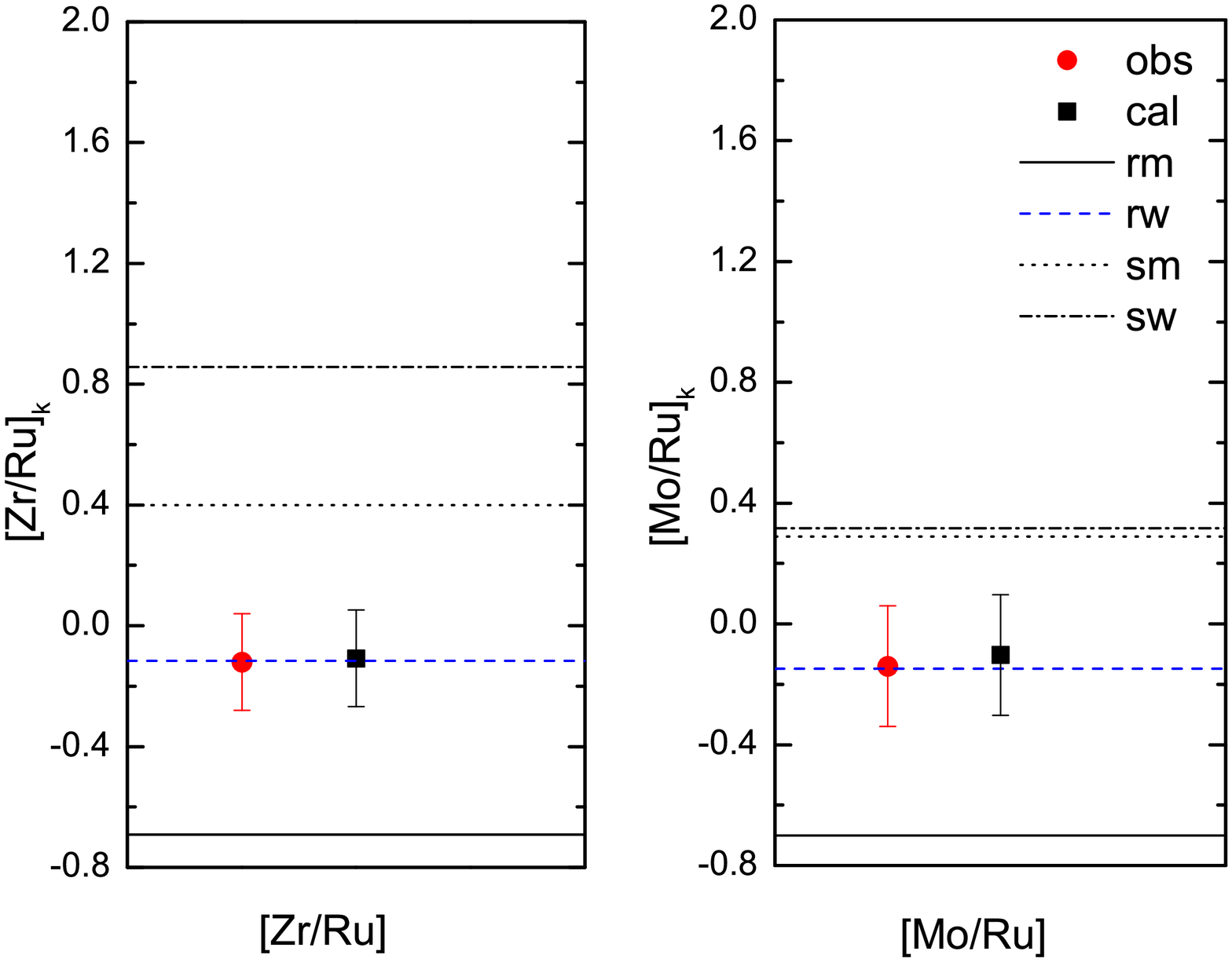}
\caption{The comparisons of the component ratios [Zr/Ru]$_k$ and [Mo/Ru]$_k$ of HD 94028. The symbols are as indicated in Figure 4.   \label{f5}}
\end{figure}

For HD 94028, the measured ratios [Zr/Fe], [Mo/Fe], and [Ru/Fe] are 0.57, 0.55, and 0.69, respectively. \citet{2016APJ...821...37R} found that the s-process models cannot reproduce the [Mo/Ru] ratio, and no combination of models can simultaneously reproduce the [As/Ge] and [Se/As] ratios and the enhanced [Zr/Fe], [Mo/Fe], and [Ru/Fe] ratios. They suggested that the i-process provides a scenario that can explain the enhancement of Mo and Ru. From Figure 3, we can see that Zr, Mo, and Ru mainly come from the weak r-process. Adopting the weak r-process abundances of Zr, Mo, and Ru calculated by \citet{2013PASP...125...143L}, the component ratios of [Zr/Ru]$_{r,w}$, [Zr/Ru]$_{r,m}$, [Mo/Ru]$_{r,w}$, and [Mo/Ru]$_{r,m}$ are about -0.12, -0.69, -0.15, and -0.7, respectively. On the other hand, the component ratios of [Zr/Ru]$_{s,m}$ and [Mo/Ru]$_{s,m}$ are about 0.40 and 0.29 \citep{2010MNRAS...404...1529B}. The ratios of [Zr/Ru]$_{s,w}$ and [Mo/Ru]$_{s,w}$  are 0.86 and 0.32 \citep{Frischknecht16}. Figure 5 shows the component ratios [Zr/Ru]$_{k}$  and [Mo/Ru]$_{k}$.  From the left panel we can see that the observed ratio [Zr/Ru] is higher than [Zr/Ru]$_{r,m}$ and lower than [Zr/Ru]$_{s,m}$ but close to the ratio [Zr/Ru]$_{r,w}$. From the right panel we can see that the observed ratio [Mo/Ru] is also close to the weak r-process ratio [Mo/Ru]$_{r,w}$. Obviously, the observed abundance ratios of HD 94028 for Zr, Mo, and Ru can be explained by the weak r-process.

\begin{figure}[h]
\includegraphics[width=3.5in,height=2.8in]{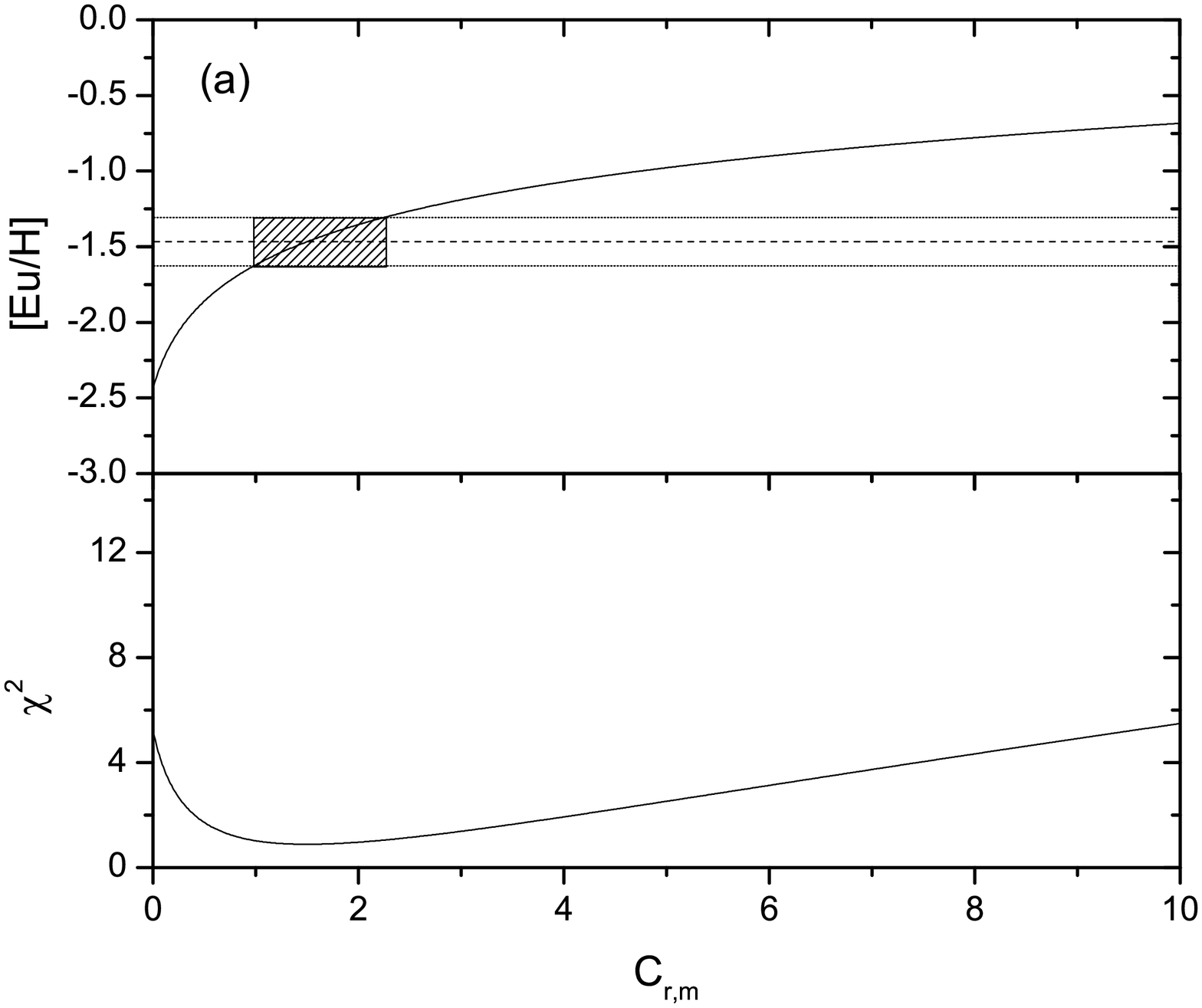}
\includegraphics[width=3.5in,height=2.8in]{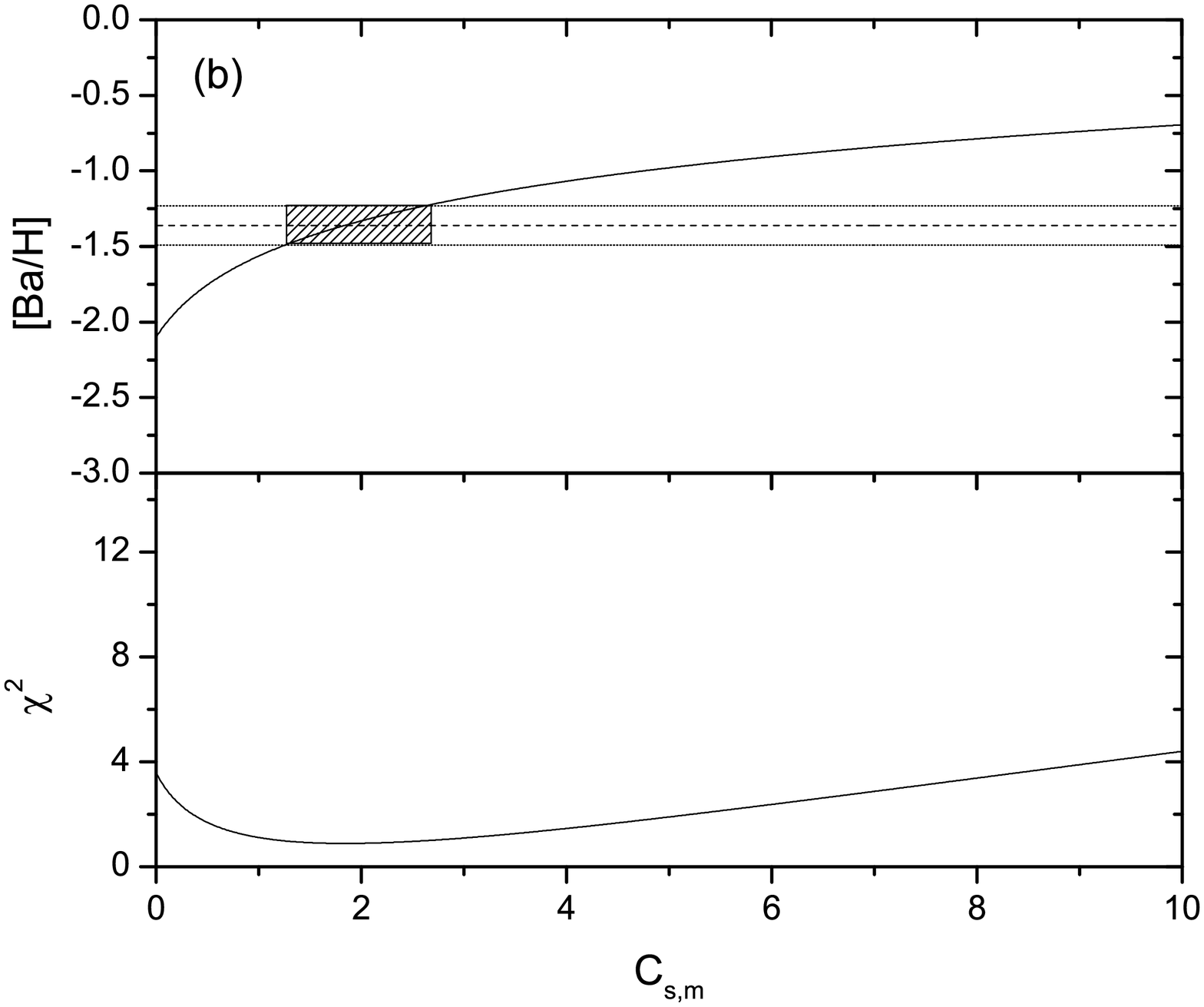}

\caption{(a) Calculated abundance ratio [Eu/H] (upper panel) and the fitting  $\chi^2$ (lower panel) as a function of component coefficient $C_{r,m}$. (b) Calculated abundance ratio [Ba/H] (upper panel) and the fitting  $\chi^2$ (lower panel) as a function of component coefficient $C_{s,m}$. Solid curves represent the calculated results. The dashed lines represent the observed values and the dotted lines refer to the observed uncertainties. The shaded regions show the allowed ranges of the component coefficients. \label{f5} }.
\end{figure}

Because the calculations in this work are based on the measured abundances of HD 94028, the observed uncertainties should be contained in the component coefficients. In order to estimate the uncertainties of the coefficients, we take the coefficients $C_{s,m}$ and $C_{r,m}$ as examples. In this case, Ba and Eu are chosen as the representative elements for the main s-process and main r-process, since the coefficient $C_{s,m}$ is most sensitive to Ba abundance and $C_{r,m}$ is most sensitive to Eu abundance. Taking $C_{pri} = 3.90$, $C_{s,m} = 1.84$, $C_{sec} = 0.17$, $C_{Ia} = 0$, we calculate the abundance ratio [Eu/H] and reduced $\chi^2$ as a function of $C_{r,m}$ and plot the ratio and $\chi^2$ in Figure 6(a). From Figure 6(a) we find that the observed ratio [Eu/H] can be accounted for in the range of $0.98 < C_{r,m} < 2.25$. Using the same method, we can derive the abundance ratio [Ba/H] and $\chi^2$ as a function of the coefficient $C_{s,m}$ and we show that in Figure 6(b). From Figure 6(b) we find that the observed ratio [Ba/H] can be accounted for in the range of $1.26< C_{s,m}< 2.62$. Similarly, the ranges of other component coefficients can be derived: $2.57 < C_{pri} < 6.48$, $0 < C_{sec} < 0.64$, and $0 < C_{Ia} < 0.34$ for star HD 94028. Because the component coefficients are constrained by the observed abundances, the uncertainties of the component coefficients are closely related to the observed uncertainties. In this case, the calculated abundance uncertainties mainly result from the observed uncertainties. For simplicity, we adopt the average uncertainty of the observed abundances as the calculated uncertainties, which are about 0.20 dex.

In order to investigate the effect of the i-process on the abundances of the lighter neutron-capture elements in HD 94028, we use $N_{i,int}$ and $C_{int}$ to represent the component abundance and coefficient of the i-process. The contribution of the i-process to the abundances of the neutron-capture elements $C_{int}N_{i,int}$ is added in Equation (1). The component abundance $N_{i,int}$ is adopted from \citet{2016APJ...821...37R} and scaled to the As abundance of the weak r-component. The calculated component coefficient of the i-process is $C_{int}=0.21$, which is about one order of magnitude smaller than that of the weak r-process. The range of the coefficient is $0<C_{int}<1.26$, in which the calculated As abundances fall within the observed limits of [As/H]. The calculated abundances of the combination of the s-, r-, and i-processes are shown in Figure 7(a). We can see that, for the elements from Ge to Te, the abundances of the i-process are lower than those of the r-process, and the origin of the lighter neutron-capture elements is mainly ascribed to the r-process. For comparison, the calculated abundances of the combination of the s- and r-processes are shown in Figure 7(b).
In Figure 7, the abundances of the weak s- and weak r-processes are included in each fit. From the figures we can see that the fitting result of the combined contributions of three processes (s-, r-, and i-processes) is close to that of the two processes (s- and r-processes) and that the contribution of the i-process has no substantial impact on the fitting result for the abundances of the lighter neutron-capture elements. The results imply that the elemental abundances of HD 94028 can be fitted by the combined contributions of the s-process and the r-process, and the additional contribution of the i-process should not be needed.

\begin{figure}[h]
\includegraphics[width=7.0in,height=2.8in]{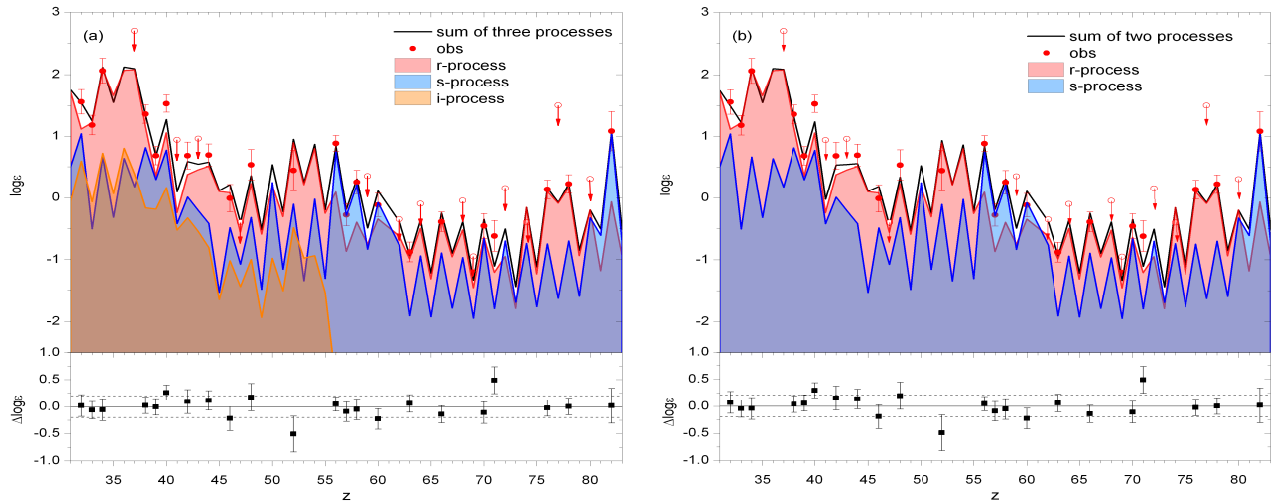}
\caption{(a)The calculated abundances of the combination of the s-process(blue), r-process(red), and i-process(orange) components. (b)The calculated abundances of the combination of the s-process(blue) and r-process(red) components. The solid black lines represent the sum of three (a) or two (b) processes. Circles with downward arrows and the filled circles represent the observed abundances, which are taken from \citet{2016APJ...821...37R}. \label{f2}}
\end{figure}

\section{Conclusions} \label{sec:conclu}
In this work, adopting the multi-component abundance method, we study the elemental abundances and explore the astrophysical origins of the elements in the metal-poor star HD 94028. The calculations show that the elements from Eu to Pt mainly come from the main r-process, confirming the conclusion about the astrophysical origin
of the heavier neutron-capture elements from Eu to Pt by \citet{2016APJ...821...37R}. Our results are listed as follows.
\begin{enumerate}
  \item The average component coefficients of the main r-process, primary process, main s-process, secondary process, and SNe Ia are 1.50, 4.0, 1.83, 0.15, and 0, respectively. The results mean that the fractions contributed by the main r-, primary, and main s-processes to the elemental abundances of HD 94028 are larger than those of the solar system and the fraction contributed by the secondary process to the abundances of this star is smaller than that of the solar system. We also find that no contribution of SNe Ia to the abundances of this star, confirming the prediction by \citet{Kobay98} that it was difficult for SNe Ia events to occur for the progenitors with [Fe/H]$\leqslant-1.0$.
  \item The lightest elements and iron group elements in the metal-poor star HD 94028 dominantly come from the primary process of the massive stars, and the contribution from the secondary process of the massive stars to these elements can be negligible.
  \item For the star HD 94028, the elements from Ge to Te mainly come from the weak r-process. However, the heavy neutron-capture elements Ba, La, Ce, Nd, and Pb dominantly originate from the metal-poor low-mass AGB star (or stars). We find that, relative to the ST case, the lower $^{13}C$-pocket efficiency (ST/18) presents a better fit to the elemental abundances of this star. The dilution factor for the s-process material of the AGB star is about 2.7\%. Adopting the observed ratio [C/Fe] as the constraint, the upper limit of the ratio [C/Fe] of the AGB star would be about 2.69.
  \item To explain the abundance features of Ge, As, and Se in HD 94028, we define the component ratios. For the ratio [As/Ge], the average component ratios of the weak r-, main r-, weak s-, and main s-processes are about 1.36, 1.31, -0.36, and -0.12, respectively. The observed ratio [As/Ge] is close to the ratio [As/Ge]$_{r,w}$. The higher ratio [As/Ge] in HD 94028 is mainly due to the higher component ratio of the weak r-process.
  \item For the ratio [Se/As], the average component ratios of the weak r-, main r-, weak s-, and main s-processes are about -0.11, 0.04, 0.01, and 0.22, respectively. The observed ratio [Se/As] is close to the component ratio [Se/As]$_{r,w}$. We find that it is only for the weak r-process that the component ratio [As/Ge] is supersolar and [Se/As] is subsolar. Obviously, the weak r-process is responsible for the observed abundance features of Ge, As, and Se in HD 94028.
  \item Because Ge is not a `pure' neutron-capture element, element As should be the starting element of the neutron-capture elements, i.e., As should be the lightest neutron-capture element. The weak r-process plays an important role in the enrichment of the light neutron-capture elements in the Galaxy. For the solar system, the fractions contributed by the weak r-process to elements As and Se are in excess of 50\%.
  \item Based on the best-fit abundances and the derived results, we can see that, in the framework of the abundance decomposition method, for HD 94028, the additional neutron-capture process (i-process) is not needed to explain the abundances observed for HD 94028 within the uncertainties.

\end{enumerate}

The results here could present new constraints on the astrophysical origins of elements for HD 94028. Obviously, more studies of the stellar abundances are required to reveal nucleosynthetic characteristics of the neutron-capture processes, including the weak r-, main r-, weak s-, and main s-processes.

\acknowledgments
We thank the referee for the insightful and constructive suggestions, which improved this paper greatly. This work has been supported by National Natural Science Foundation of China under grants 11673007, 11273011, 11403007, 11547041, 11643007, and 11773009, the Natural Science Foundation of Hebei Province under grant A2011205102, the Program for Excellent Innovative Talents in University of Hebei Province under grant CPRC034, and the Innovation Fund Designated for Graduate Students of Hebei Province under grant SJ2016023.


\begin{thebibliography}{}
\bibitem[Aoki et al. (2001)]{Aoki01} Aoki, W., Ryan, S.~G., Norris, J.~E., Beers, T.~C., \& Ando, H., et al. 2001, \apj, 561, 346
\bibitem[Aoki et al. (2002)]{Aoki02} Aoki, W., Ryan, S.~G., Norris, J.~E., et al. 2002, \apj, 580, 1149
\bibitem[Arcones \& Bliss (2014)]{Arcones14} Arcones, A., \& Bliss, J. 2014, J.Phys.G:Nucl.part.Phys., 41, 044005
\bibitem[Arlandini et al.(1999)]{Arlan99} Arlandini, C., K\"{a}ppeler, F., Wisshak, K., et al. 1999, \apj, 525, 886
\bibitem[Arnould \& Goriely (2003)]{Arnould03} Arnould, M., \& Goriely, S. 2003, PhR, 384, 1
\bibitem[Bisterzo et al. (2010)]{2010MNRAS...404...1529B} Bisterzo, S., Gallino, R., Straniero, O., et al. 2010, \mnras, 404, 1529
\bibitem[Bisterzo et al. (2011)]{Bisterzo11} Bisterzo, S., Gallino, R., Straniero, O., et al. 2011, \mnras, 418, 284
\bibitem[Bisterzo et al. (2017)]{Bisterzo17} Bisterzo, S., Travaglio, C., Wiescher, M., K\"{a}ppeler, F., \& Gallino, R. 2017, \apj, 835, 97B
\bibitem[Burris et al. (2000)]{Burris00} Burris D.~L., Pilachowski, C.~A., Armandroff, T.~E., et al. 2000, \apj, 544, 302
\bibitem[Busso et al.(1999)]{Busso99} Busso, M., Gallino, R., \& Wasserburg, G.~J. 1999, ARA\&A, 37, 239
\bibitem[Busso et al.(2001)]{Busso01} Busso, M., Gallino, R., Lambert, D.~L. et al.  2001, \apj, 557, 802
\bibitem[Campbell et al. (2010)]{Camp10} Campbell, S.~W., Lugaro, M., Karakas, A.~I. 2010, A\&A, 522, L6
\bibitem[Cohen et al. (2003)]{Cohen03} Cohen, J.~G., Christlieb, N., Qian, Y.-Z., Wasserburg, G.~J., 2003, \apj, 588, 1082
\bibitem[Cowan et al. (2005)]{Cowan05} Cowan, J.~J., Sneden, C., Beers, T.~C., et al. 2005, \apj, 627, 238
\bibitem[Eichler et al. (1989)]{Eichler89} Eichler, D., Livio, M., Piran, T., \& Schramm, D. N. 1989, Natur, 340, 126
\bibitem[Freiburghaus et al. (1999)]{Freibur99} Freiburghaus, C., Rosswog, S., Thielemann, F.-K. 1999, \apjl, 525, 121
\bibitem[Frischknecht et al. (2012)]{Frischknecht12} Frischknecht, U., Hirschi, R., Thielemann, F.-K. 2012, A\&A, 538, L2
\bibitem[Frischknecht et al. (2016)]{Frischknecht16} Frischknecht, U., Hirschi, R., Pignatari, M. 2016, \mnras, 456, 1803
\bibitem[Gallino et al. (1998)]{Galli98} Gallino, R., Arlandini, C., Busso, M., et al. 1998, \apj, 497, 388
\bibitem[Hansen et al. (2014)]{Hansen14} Hansen, C. J., Montes, F., \& Arcones, A. 2014, ApJ, 797, 123H
\bibitem[Herwig et al. (2011)]{Herw11} Herwig, F., Pignatari, M., Woodward, P.~R., et al.  2011, \apj, 727, 89
\bibitem[Honda et al. (2006)]{Honda06} Honda, S., Aoki, W., Ishimaru, Y., Wanajo, S., \& Ryan, S.~G. 2006, \apj, 643, 1180
\bibitem[Honda et al. (2007)]{Honda07} Honda, S., Aoki, W., Ishimaru, Y., \& Wanajo, S. 2007, \apj, 666, 1189
\bibitem[Ishimaru et al. (2005)]{Ishimaru05} Ishimaru, Y., Wanajo, S., Aoki, W., et al. 2005, NuPhA, 758, 603
\bibitem[Ji et al. (2016)]{Ji2016} Ji, A.~P., Frebel, A., Chiti, A., \& Simon, J.~D. 2016, Natur, 531, 610
\bibitem[Johnson \& Bolte (2002)]{Johnson02} Johnson, J.~A., \& Bolte, M. 2002, \apjl, 579, L87
\bibitem[Johnson \& Bolte (2004)]{04} Johnson, J.~A., \& Bolte, M. 2004, \apj, 605, 462
\bibitem[K\"{a}ppeler et al. (2011)]{Kappeler11} K\"{a}ppeler, F., Gallino, R., Bisterzo, S., \& Aoki, W. 2011, RvMP, 83, 157
\bibitem[Karakas \& Lattanzio (2014)]{Karakas14} Karakas, A.~I., \& Lattanzio, J.~C., 2014, PASA, 31, e030
\bibitem[Kobayashi et al. (1998)]{Kobay98} Kobayashi, C., Tsujimoto T., Nomoto, K., et al.  1998, \apj, 503, L155
\bibitem[Kobayashi et al. (2006)]{2006APJ...653...1145K} Kobayashi, C., Umeda, H., Nomoto, K., et al.  2006, \apj, 653, 1145
\bibitem[Komiya et al. (2014)]{Komiya14} Komiya, Y., Yamada, S., Suda, T., \& Fujimoto, M. Y. 2014, \apj, 783, 132
\bibitem[Komiya \& Shigeyama (2016)]{Komiya16} Komiya, Y., \& Shigeyama, T. 2016, \apj, 830, 76
\bibitem[Lattimer \& Schramm (1974)]{Lattimer74} Lattimer, J. M., \& Schramm, D. N. 1974, \apjl, 192, L145
\bibitem[Li et al. (2013a)] {2013PASP...125...143L} Li, H.-J., Shen, X.-J., Liang, Sh., Cui, W.-Y., Zhang, B.  2013a, \pasp, 125, 143
\bibitem[Li et al. (2013b)] {Li2013} Li, H.-J., Cui, W.-Y., Zhang, B.  2013b, \apj, 775, 12
\bibitem[Lugaro et al. (2015)]{Lugaro2015} Lugaro, M., Campbell, S.~W., Van Winckel, H., et al. 2015, A\&A, 583, A77
\bibitem[Lucatello et al. (2003)]{Lucatello03} Lucatello, S., Gratton, R., Cohen, J.~G., et al. 2003, \aj, 125, 875
\bibitem[Masseron et al. (2010)]{Masse10} Masseron,T., Johnson, J.~A., Plez, B., et al.  2010, A\&A, 509, A93
\bibitem[Mathews et al. (1992)]{Mathews92} Mathews, G.~J., Bazan, G., \& Cowan, J.~J. 1992, \apj, 391, 719
\bibitem[Matteucci et al. (2014)]{Matteucci14} Matteucci, F., Romano,D., Arcones, A., et al. 2014, \mnras, 438, 2177
\bibitem[Montes et al. (2007)]{Montes07} Montes, F., Beers, T.~C., Cowan, J., et al. 2007, \apj, 671, 1685
\bibitem[Niu et al. (2014)]{2014MNRAS...443...2426N} Niu, P., Liu, W.-L., Cui, W.-Y., Zhang, B.  2014, \mnras, 443, 2426
\bibitem[Niu et al. (2015)]{Niu15} Niu, P., Cui, W.-Y., Zhang, B.  2015, \apj, 813, 56
\bibitem[Ohkubo T. et al. (2006)]{Ohkubo06} Ohkubo, T., Umeda, H., Nomoto, K., \& Yoshida T. 2006, AIP Conf.Proc. 847, 458
\bibitem[Peterson (2011)]{2011APJ...742...21P} Peterson, R. C. 2011, \apj, 742, 21
\bibitem[Pignatari et al. (2010)]{Pignatari10} Pignatari, M., Gallino, R., Heil, M., et al. 2010, \apj, 710, 1557
\bibitem[Pignatari et al. (2016)]{Pignatari16} Pignatari, M., Herwig, F., Hirschi, R., et al. 2016, arXiv:1307.6961v3
\bibitem[Pumo (2012)]{Pumo12} Pumo M.~L. 2012, arXiv:1202.6577v2
\bibitem[Qian \& Wasserburg (2001)]{Qian01} Qian, Y.-Z., \& Wasserburg, G.~J. 2001, \apj, 559, 925
\bibitem[Qian \& Wasserburg (2007)]{Qian07} Qian, Y.-Z., \& Wasserburg, G.~J. 2007, PhR, 442, 237
\bibitem[Ramirez-Ruiz et al. (2015)]{Ramirez15} Ramirez-Ruiz, E., Trenti, M., MacLeod, M., et al. 2015, \apjl, 802, L22
\bibitem[Raiteri et al. (1993)]{Raiteri93} Raiteri, C.~M., Gallino, R., Busso, M., Neuberger, D., \& K\"{a}ppeler, F. 1993, \apj, 419, 207
\bibitem[Rayet et al. (1990)]{Rayet90} Rayet, M., Prantzos, N., \& Arnould, M. 1990, A\&A, 227, 271
\bibitem[Roederer et al. (2010)]{2010APJ...714...L123R} Roederer, I.~U., Sneden, C., Lawler, J.~E., Cowan, J.~J. 2010, \apjl, 714, L123
\bibitem[Roederer (2012)]{2012APJ...756...36R} Roederer, I.~U. 2012, \apj, 756, 36
\bibitem[Roederer et al. (2014)]{2014AJ...147...136R} Roederer, I.~U., Preston, G.~W., Thompson, I.~B., et al. 2014, \aj, 147, 136
\bibitem[Roederer et al. (2016)]{2016APJ...821...37R} Roederer, I.~U., Karakas, A.~I., Pignatari, M., Herwig, F.  2016, \apj, 821, 37
\bibitem[Sivarani et al. (2004)]{Sivarani04} Sivarani, T., et al., 2004, A\&A, 413, 1073
\bibitem[Sneden et al. (2008)]{Sned08} Sneden, C., Cowan, J.~J, \& Gallino, R. 2008, ARA\&A, 46, 261
\bibitem[Straniero et al. (2006)]{Straniero06} Straniero O., Gallino R., Cristallo S., 2006, NuPhA, 777, 17
\bibitem[Thielemann et al. (2011)]{Thielemann11} Thielemann, F.-K., Arconesa, A., K\"{a}ppeli, R., et al. 2011, PrPNP, 66, 346
\bibitem[Timmers et al. (1995)]{1995APJ...98...617T} Timmes, F.~X., Woosley, S.~E., \& Weaver, T.~A. 1995, \apjs, 98, 617
\bibitem[Travaglio et al. (1999)]{Trava99} Travaglio, C., Galli, D., Gallino, R., et al.  1999, \apj, 521, 691
\bibitem[Travaglio et al. (2004)]{Trava04} Travaglio, C., Gallino, R., Arnone, E., et al.  2004, \apj, 601, 864
\bibitem[Tsujimoto \& Shigeyama (2014)]{Tsujimoto14} Tsujimoto, T., \& Shigeyama, T. 2014, ApJL, 795, L18
\bibitem[Van Eck et al. (2001)]{Van01} Van Eck, S., Goriely, S., Jorissen, A., Plez, B., 2001, Natur, 412, 793
\bibitem[Van Eck et al. (2003)]{Van03} Van Eck, S. ,Goriely, S., Jorissen, A., Plez, B., 2003, A\&A, 404, 291
\bibitem[Wheeler et al. (1998)]{Whee98} Wheeler, J.~C., Cowan, J.~J., Hillebrandt, W.  1998, \apjl, 493, L101
\bibitem[Woosley et al.(2002)]{Woos02} Woosley, S.~E.,Heger, A., \& Weaver, T.~A.  2002, RvMP, 74, 1015

\end{thebibliography}
\end{document}